\documentclass[lettersize,journal]{IEEEtran}
\usepackage{amsmath,amsfonts}
\usepackage{algorithmic}
\usepackage{algorithm}
\usepackage{array}
\usepackage[caption=false]{subfig}
\usepackage{textcomp}
\usepackage{stfloats}
\usepackage{url}
\usepackage{verbatim}
\usepackage{graphicx}
\usepackage[noadjust]{cite}
\usepackage{dsfont}
\usepackage{footmisc}
\usepackage{color}
\usepackage{flushend}

\begin{document}

\title{\textcolor{black}{Increasing Data Rate through Shaping on} Wireless\\Channels Subject to Mobility and Delay Spread}

\author{Sandesh Rao Mattu$^*$, Nishant Mehrotra$^*$, Robert Calderbank~\IEEEmembership{Life Fellow,~IEEE}
\thanks{This work is supported by the National Science Foundation under grants 2342690 and 2148212, in part by funds from federal agency and industry partners as specified in the Resilient \& Intelligent NextG Systems (RINGS) program, and in part by the Air Force Office of Scientific Research under grants FA 8750-20-2-0504 and FA 9550-23-1-0249. \\
The authors are with the Department of Electrical and Computer Engineering, Duke University, Durham, NC, 27708, USA (email: \{sandesh.mattu,~nishant.mehrotra,~robert.calderbank\}@duke.edu). \\
$*$ denotes equal contribution.}
}



\maketitle

\begin{abstract}
This letter describes how to improve performance of cellular systems by combining non-equiprobable signaling (shaping) with low-density parity check (LDPC) coding for an orthogonal frequency division multiplexing system. 
We focus on improving performance at the cell edge, where the 5G standard specifies a suite of LDPC codes with different rates that are applied to 4-QAM. We employ the method of shaping on rings which adds to the transmission rate as it shapes the input distribution. We double the size of the $4$-QAM constellation by introducing a second shell of signal points, and we implement non-equiprobable signaling through a shaping code which selects the high energy shell less frequently than the low energy shell. 
We describe how to combine coding and shaping by integrating shaping into the calculation of log-likelihood ratios (LLRs) necessary for decoding LDPC codes. 
We employ rate $1/2$ LDPC coding and select the rate of the shaping code to match that of rate $3/4$ LDPC coding using $4$-QAM. 
We present simulation results for a representative Veh-A channel showing gains of $4$ dB at a bit error rate (BER) of $10^{-3}$. 
When we choose an LDPC code from the 5G suite to match the BER performance of rate $1/2$ LDPC coding with shaping we show that transmission rate can be improved by $20 $\%. 
\end{abstract}

\begin{IEEEkeywords}
OFDM, shaping, non-equiprobable signaling, LDPC.
\end{IEEEkeywords}

\section{Introduction}
\label{sec:intro}

In his 1948 paper, Shannon~\cite{shannon1948a} recognized that signals input to a Gaussian channel should themselves be selected with a Gaussian distribution. Wireless channels subject to mobility and delay spread are not Gaussian channels, but it is still advantageous to select signals with large energy less frequently than signals with small energy. This can be achieved by geometric shaping (GS), where a standard QAM constellation is warped so that points close to the origin are closer together and points closer to the periphery are further apart. A second method is probabilistic shaping (PS), where the constellation points are kept as is, and the distribution over them is changed. These shaping methods apply to any coding and modulation scheme, they do not introduce additional rate, and they enable modest improvements in performance (typically less than 1dB)~\cite{3GPP_R1_2508738}. 

In their 1998 survey paper, Forney and Ungerboeck~\cite{forney1998modulation} described how coding and shaping can be combined to allow capacity to be approached on any linear Gaussian channel. A particular focus is the 1994/1996 ITU-T V.34 standard that enabled wireline access to the internet at rates up to $33.6$ kb/s. The survey describes how error performance on the Gaussian channel is determined by the figure of merit $d^2/P$ where $P$ is the average transmitted signal power and $d^2$ is the minimum squared distance of the code. We can transmit an extra bit by doubling the size of a baseline QAM constellation and signaling equiprobably. However, this doubles the average transmitted signal power $P$, and to maintain error performance we would need to double the minimum squared distance $d^2$. In the 1990s computing power was very limited and the complexity of decoding the more powerful code was typically out of reach. The V.34 standard increased transmission rate without increasing coding complexity by using a rate $R$ shaping code to transmit additional information. The V.34 standard includes the method of shaping on rings proposed by Calderbank and Ozarow~\cite{calderbank2002nonequiprobable} where a QAM constellation is partitioned into annular subconstellations of equal size by scaling a basic circular region, and the shaping code selects the subconstellations with different probabilities. Equiprobable signaling would match the rate by scaling the baseline constellation by $2^R$, thereby increasing average transmitted signal power by the same factor. When the power required by non-equiprobable signaling is less than $2^RP$ there is a shaping gain. It is possible to closely approach the asymptotic shaping gain of $1.53$ dB on a Gaussian channel by shaping on $8$ annuli (see~\cite{calderbank2002nonequiprobable} for details). 

\textcolor{black}{We use the method of shaping on rings to improve 5GNR performance at the cell edge. The 5GNR standard specifies rate $1/2$ and rate $3/4$ LDPC codes which use 4-QAM at the cell edge. At a bit error rate of $10^{-3}$ the gap between the performance of the rate $1/2$ code and the rate $3/4$ code is significant - approximately $4$ dB. We bridge this gap by combining the rate $1/2$ LDPC code with a shaping code. We expand the baseline $4$-QAM constellation to one of the $8$-QAM constellations\textcolor{black}{\footnote{\textcolor{black}{We refer to both constellations as 8-QAM since each has 8 signal points, and we have observed no difference in performance between the two constellations.}\label{foot:16_qam}}} shown in Fig.~\ref{fig:const_4_qam} by introducing a second shell of signal points, and we employ shaping codes that select the high energy shell less frequently than the low energy shell. The rate $3/4$ LDPC code transmits $1.5$ bits per QAM symbol, and we match this rate by introducing a shaping code with rate $0.5$ bits per QAM symbol. We ask the fundamental question of whether it is better to increase rate at the cell edge in 5GNR by introducing shaping or by increasing the rate of the LDPC code. We present simulation results for a representative Veh-A channel showing that rate $1/2$ LDPC coding with shaping improves on rate $3/4$ LDPC coding by $4$ dB at a bit error rate (BER) of $10^{-3}$. Our results demonstrate that at the cell edge where transmission rates are low, there is significant value in using shaping to contribute transmission rate. This does not appear possible with the GS and PS methods currently under consideration at 3GPP. }



\textcolor{black}{In order to combine coding and shaping it is necessary to integrate shaping into the calculation of the log-likelihood ratios (LLRs) necessary for decoding LDPC codes. We consider the orthogonal frequency division multiplexing (OFDM) system model because that is the current focus of 3GPP, but the method of shaping on rings can be applied to any modulation, for example to orthogonal time frequency space (OTFS). Section~\ref{sec:shaping} presents our algorithm for calculating LLRs. A feature of this algorithm that may be of independent interest is the way it takes signal energy into account. The gains from our method of shaping diminish as the transmission rate increases, but we are still able to present an example in Section~\ref{subsec:16_qam_23_shap} showing significant gains for $16$-QAM.}

\textit{Notation:} $x$ denotes a complex scalar, $\mathbf{x}$ denotes a vector with $n$th entry $\mathbf{x}[n]$, and $\mathbf{X}$ denotes a matrix with $(n,m)$th entry $\mathbf{X}[n,m]$. $(\cdot)^{\top}$ denotes transpose, and 
$(\cdot)^{\mathsf{H}}$ denotes complex conjugate transpose. $\mathbb{C}$, $\mathbb{R}$, $\mathbb{Z}$, $\mathbb{R}_+$, and $\mathbb{Z}_+$ respectively denote the set of complex numbers, real numbers, integers, positive real numbers, and  positive integers. $\star$ denotes the convolution operation. $\mathbb{E}$ denotes the expectation operator. $\mathsf{vec}(\cdot)$ denotes the column-wise vectorization operation and $\mathsf{diag}(\cdot)$ of a matrix returns a vector of its diagonal entries, $\Vert\cdot\Vert$ denotes the $2$-norm of a vector, and $\vert\cdot\vert$ denotes the cardinality of a set or absolute value of a complex number. $\otimes$ denotes the Kronecker product of matrices. $\mathcal{CN}(a, b)$ denotes circularly symmetric complex Gaussian random variable with mean $a$ and variance $b$. $\mathsf{U}[a, b)$ denotes a uniform random variable with limits $a$ (inclusive) and $b$ (exclusive).



\vspace{-1mm}
\section{System Model}
\label{sec:sys_model}
Consider an OFDM frame with $M$ subcarriers and $N$ symbols. Let $\Delta f$ be the subcarrier spacing. The OFDM frame occupies bandwidth $B=M\Delta f$ and time $T = N/\Delta f$. Each OFDM symbols occupies time $T_\mathsf{s} = 1/\Delta f$. The system model can be represented as (see~\cite{mattu2026increasing} for a detailed derivation of the system model):
Let $\mathbf{X}\in \mathbb{C}^{M\times N}$ denote the matrix of information symbols. At the transmitter, the information symbols are converted to time domain as:
\begin{align}
    \label{eq:sys_model1}
    \mathbf{x} = \mathsf{vec}(\mathbf{F}_M^{\mathsf{H}}\mathbf{X}) = (\mathbf{I}\otimes\mathbf{F}_M^{\mathsf{H}})\mathsf{vec}(\mathbf{X}),
\end{align}
where $\mathbf{x} \in \mathbb{C}^{MN \times 1}$ is the time-domain vector corresponding to the information symbols and $\mathbf{F}_M$ is the $M$-point discrete Fourier transform (DFT) matrix\footnote{In OFDM systems, a cyclic prefix is required for avoiding interference from neighboring OFDM symbols. We assume that it is added at the transmitter in time domain and removed at the receiver in the time domain and does not influence the system model.}. The time domain signal is mounted on a pulse $p(t)$:
\begin{align}
    \label{eq:sys_model2}
    s(t) = \sum_{n=0}^{MN-1}\mathbf{x}[n]p(t-nT_0),
\end{align}
where $T_0 = 1/B$ is the sampling interval. The time domain signal is transmitted through a doubly-selective channel whose representation in the delay-Doppler domain given by:
\begin{align}
    \label{eq:sys_model3}
    h(\tau, \nu) = \sum_{p=0}^{P-1}h_p\delta(\tau-\tau_p)\delta(\nu-\nu_p),
\end{align}
where $h_p, \tau_p, \nu_p$ respectively denote the channel gain, the delay, and the Doppler of the $p$th path of the $P$ path channel. The received time domain signal can be represented as:
\begin{align}
    \label{eq:sys_model4}
    r(t) &= \int_{\tau}\int_{\nu}h(\tau, \nu)s(t-\tau)e^{j2\pi\nu(t-\tau)}d\tau d\nu + n(t) \nonumber \\
    &= \sum_{p=0}^{P-1}h_ps(t-\tau_p)e^{j2\pi\nu_p(t-\tau_p)} + n(t),
\end{align}
where $n(t)$ is the additive while Gaussian noise with distribution $\mathcal{CN}(0, N_0)$. The delay and the Doppler of each path can be expressed in terms of the corresponding indices as:
\begin{align}
    \label{eq:sys_model5}
    \tau_p = \frac{k_p}{B}, \nu_p = \frac{l_p}{T},
\end{align}
where $k_p \in \mathbb{R}_+$ and $l_p\in \mathbb{R}$ are the delay and Doppler indices, respectively, corresponding to each channel path. Note that since both $k_p$ and $l_p$ can take real values (and not limited to integers), the channel model has \textit{fractional} delay-Doppler indices, which is practical. At the receiver, the received time domain signal is passed through a pulse $p^\ast(-t)$ matched to the transmit pulse:
\begin{align}
    \label{eq:sys_model6}
    y(t) 
    &= \sum_{p=0}^{P-1}\sum_{n=0}^{MN-1}h_pe^{j\frac{2\pi}{MN}l_pn}\mathbf{x}[n]g(t-(k_p+n)T_0) + w(t),
\end{align}
where $g(t) = p(t)e^{j2\pi\nu_p t}\star p^\ast(-t)$, and $w(t) = n(t) \star p^\ast(-t)$. Sampling~\eqref{eq:sys_model6} at $t = mT_0, m = 0, 1, \cdots, MN-1$:
\begin{align}
    \label{eq:sys_model7}
    \mathbf{y}[m] = \sum_{p=0}^{P-1}\sum_{n=0}^{MN-1}h_pe^{j\frac{2\pi}{MN}l_pn}\mathbf{x}[n]\mathbf{g}[m-(k_p+n)] + \mathbf{w}[m],
\end{align}
which can be expressed in matrix vector form as:
\begin{align}
    \label{eq:sys_model8}
    \mathbf{y} = \mathbf{H}\mathbf{x} + \mathbf{w},
\end{align}
where $\mathbf{y}, \mathbf{x}, \mathbf{n} \in \mathbb{C}^{MN\times 1}$ are the received, transmitted, and noise vector, respectively, and $\mathbf{H}\in \mathbb{C}^{MN \times MN}$ is the effective channel matrix which encompasses the effect of pulse shaping and the physical channel. The $m$th row and $n$th column of the channel matrix is:
\begin{align}
    \label{eq:sys_model9}
    \mathbf{H}[m, n] = \sum_{p=0}^{P-1}h_pe^{j\frac{2\pi}{MN}l_pn}\mathbf{g}[m-(k_p+n)].
\end{align}
Finally, the time domain system model in~\eqref{eq:sys_model8} can be represented in the frequency domain as:
\begin{align}
    \label{eq:sys_model10}
    \mathbf{y}_{\mathsf{F}} = \mathbf{H}_{\mathsf{F}}\mathbf{x}_{\mathsf{F}} + \mathbf{w}_{\mathsf{F}}.
\end{align}
where $\mathbf{y}_{\mathsf{F}} = (\mathbf{I}\otimes \mathbf{F}_M)\mathbf{y}$, $\mathbf{x}_{\mathsf{F}} = \mathsf{vec}(\mathbf{X})$, and $\mathbf{w}_{\mathsf{F}} = (\mathbf{I}\otimes \mathbf{F}_M)\mathbf{w}$.

\subsection{Channel estimation}
\label{subsec:chan_est}
In practice, systems are required to estimate the channel at the receiver. 
To estimate the channel, a reference symbol or pilot is transmitted. In this paper, we consider the Kronecker pilot pattern (also called the Type-2 demodulation reference signal (DMRS))~\cite{3gpp_ts_38_211, 3gpp_ts_38_212}. In this setting pilots are placed on specific subcarriers in an OFDM symbol. The estimates at the pilot locations are obtained using a least squares estimator~\cite{nee_prasad_ofdm_2000} and to get the estimate at the rest of the locations a two-dimensional interpolation is carried out. For the two-dimensional interpolation, the (pre-computed) time and frequency covariance matrices are used\footnote{Here, we consider the diagonal values when we compute the covariance matrices. This is because under fractional delay-Doppler channel, the channel matrix is not strictly diagonal due to inter-carrier interference (ICI) in OFDM. Taking the whole matrix introduces ICI effects into the time and frequency covariance matrix.}, which are computed respectively as:
\begin{align}
    \label{eq:sys_model11}
    \mathbf{C}_\mathsf{t} = \mathbb{E}[\mathsf{diag}(\mathbf{H}_{\mathsf{F}}^{\mathsf{H}})\mathsf{diag}(\mathbf{H}_{\mathsf{F}})], \mathbf{C}_\mathsf{f} = \mathbb{E}[\mathsf{diag}(\mathbf{H}_{\mathsf{F}})\mathsf{diag}(\mathbf{H}_{\mathsf{F}}^{\mathsf{H}})].
\end{align}
A linear minimum mean squared error (LMMSE) interpolator is used for the two-dimensional interpolation of the channel values~\cite{hoeher1997two-dimensional} which gives the estimate of the diagonals of $\mathbf{H}_{\mathsf{F}}$. This forms the estimated channel matrix which is used for equalization.


\section{Shaping}
\label{sec:shaping}

\begin{figure}
    \centering
    \includegraphics[width=\linewidth]{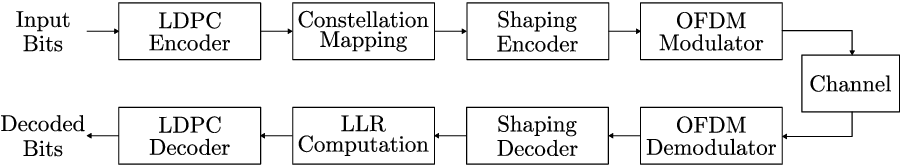}
    \caption{Block diagram of the OFDM communication scheme with shaping.}
    \label{fig:blk_dia}
\end{figure}
Figure~\ref{fig:blk_dia} shows the block diagram of the overall communication scheme. At the transmitter, information bits are encoded through a low-density parity check (LDPC) encoder, followed by mapping to constellation symbols (4-QAM, for example). The symbols are then passed through a shaping encoder. The resulting constellation symbols are modulated using OFDM and transmitted over a doubly-selective channel. At the receiver, the received symbols are OFDM demodulated. This includes channel estimation and equalization. This is followed by a shaping decoder. The OFDM demodulated symbols are passed to the LLR computation block and then finally to the LDPC decoder to get the decoded information bits. In the following, we describe the shaping encoder and decoder in detail.

\subsection{Shaping Encoder}
\label{subsec:shap_enc}
\begin{figure}
    \centering
    \includegraphics[width=\linewidth]{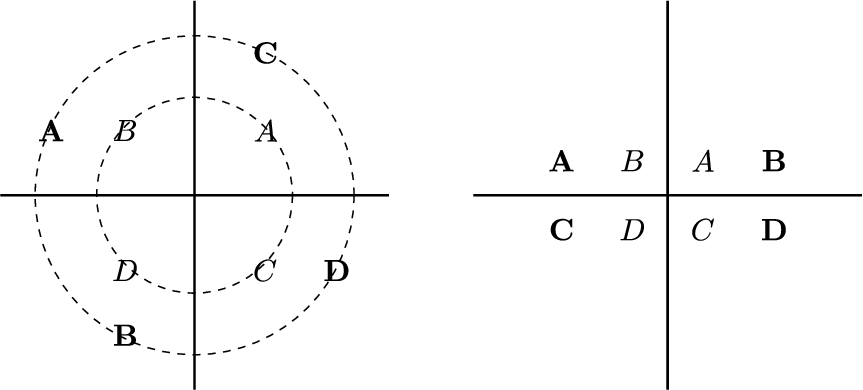}
    \caption{\textcolor{black}{Two $8$-QAM constellations with identical performance.}}
    \label{fig:const_4_qam}
\end{figure}
We augment the original $4$-QAM constellation to obtain the $8$-QAM constellations shown in Fig.~\ref{fig:const_4_qam}. Points in the augmented constellation are labeled $\mathbf{A}$, $\mathbf{B}$, $\mathbf{C}$, or $\mathbf{D}$ with different labels corresponding to different residue classes modulo $4$. Every label appears once in the inner constellation and once in the outer constellation.
The choice between the inner and outer constellation is provided by the shaping code. Note that the shaping code is not actually transmitted, the extra information bits are conveyed through power during transmission.

\begin{figure}
    \centering
    \includegraphics[width=\linewidth]{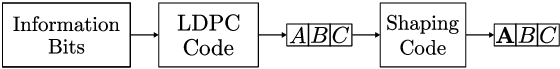}
    \caption{
    The shaping codeword $[1\ 0\ 0]$ transforms a sequence $[A\ B\ C]$ of labels to a sequence $[\mathbf{A}\ B\ C]$ of transmitted signal points.}
    \label{fig:shap_enc_block_dia}
\end{figure}

An $l$-sparse shaping code of length $z$ is a collection of binary vectors with at most $l$ entries equal to $1$. The LDPC code selects a sequence of $z$ $4$-QAM points (labels). The labels are input to the shaping code which chooses between the representatives in the inner and outer constellations. If $f$ is the function that maps an inner constellation point to the corresponding outer point, then the shaping code performs the following function.
\begin{align}
    \label{eq:shap_enc1}
    \bar{\mathbf{s}}[i] = \begin{cases}
        \mathbf{s}[i] \quad\quad~ \mathrm{if} \ \mathbf{c}[i] = 0\\
        f(\mathbf{s}[i]) \quad \mathrm{if} \ \mathbf{c}[i] = 1
    \end{cases},
\end{align}
for $i=0, 1, \cdots, z-1$ and $\bar{\mathbf{s}}\in \mathbb{C}^{z\times 1}$ is the encoded $z$-length vector that is transmitted. The encoder transmits information symbols from the inner constellation whenever the shaping code at the corresponding index is $0$ and from the outer constellation otherwise. Note that, the all $0$ shaping codeword corresponds to no shaping (see Fig.~\ref{fig:shap_enc_block_dia}).


These constellation symbols are OFDM modulated and transmitted over a channel. At the receiver, the channel estimation and equalization are carried out. The equalized symbols are passed to a shaping decoder, the details of which are presented below.

\begin{algorithm}
    \caption{Compute LLRs from equalized OFDM symbols with shaping.}
    \label{alg:llrs}
    \begin{algorithmic}[1]
        \STATE \textbf{Inputs:}
          Equalized received symbols $\mathbf{y} \in \mathbb{C}^{n_{\mathsf{sym}}\times 1}$ , constellation points $\mathcal{S}$ with $q = \vert\mathcal{S}\vert$, $\mathcal{S}_{\mathsf{i}}, \mathcal{S}_{\mathsf{o}}$, with $l=\vert\mathcal{S}_{\mathsf{i}}\vert = \vert\mathcal{S}_{\mathsf{o}}\vert$, label map $\ell:\{0,\cdots,q-1\}\to\{0,\dots,l-1\}$, noise variance $N_0>0$, log-priors $\log \boldsymbol{\pi}_i$ for $i=0,\cdots,q-1$, and small constant $\varepsilon>0$ for numerical stability.
        \STATE \textbf{Set} $n_b = \lceil\log_2 l\rceil, \mathbf{u} = \mathbf{d} = \mathbf{0}_{q\times 1}, \mathbf{\Lambda} = \mathbf{0}_{l\times 1}, \mathrm{LLR} = []$
        \FOR{$s = 0$ \textbf{to} $n_{\mathsf{sym}}-1$}
          \STATE \textbf{Set} $y = \mathbf{y}[s]$
          %
          \FOR{$i = 0$ \textbf{to} $q-1$}
            \STATE $\mathbf{d}[i] = \Vert y - \mathcal{S}_i \Vert^2$, $\mathcal{S}_i$ is the $i$th symbol in $\mathcal{S}$
            \STATE $\mathbf{u}[i] = -\frac{d_i}{N_0}$
            \STATE $\mathbf{u}_i = \mathbf{u}_i + \log \boldsymbol{\pi}_i$
          \ENDFOR
          \FOR{$\ell_{\mathsf{val}} = 0$ \textbf{to} $l-1$}
            \STATE $\mathcal{T}_\ell = \{ i \mid \ell(i)=\ell_{\mathsf{val}} \}$
            \STATE $m_\ell = \underset{i \in \mathcal{T}_\ell}{\max} \  \mathbf{u}[i]$
            \STATE $s_\ell = \sum_{i\in \mathcal{T}_\ell} \exp\big(\mathbf{u}[i] - m_\ell\big)$
            \STATE $\mathbf{\Lambda}[\ell] = m_\ell + \log\big(s_\ell + \varepsilon\big)$
          \ENDFOR
          \STATE $\mathbf{l}_{\mathsf{sym}} = \mathbf{0}_{n_b\times 1}$
          \FOR{$b= 0$ \textbf{to} $n_b-1$}
            \STATE Define $\mathcal{L}_0^{(b)} = \{\ell : \text{bit } b \text{ of label } \ell = 0\}, \mathcal{L}_1^{(b)} = \{\ell : \text{bit } b \text{ of label } \ell = 1\}.
            $
            \STATE $m_0 = \underset{k\in\mathcal{L}_0^{(b)}}{\max} \mathbf{\Lambda}[k]$, $m_1 = \underset{k\in\mathcal{L}_1^{(b)}}{\max} \mathbf{\Lambda}[k]$
            \STATE $s_0 = \sum_{\ell\in\mathcal{L}_0^{(b)}} e^{(\mathbf{\Lambda}[\ell] - m_0)}, s_1 =\sum_{\ell\in\mathcal{L}_1^{(b)}} e^{(\mathbf{\Lambda}[\ell] - m_1)}$
            \STATE $\mathbf{l}_{\mathsf{sym}}[b] = \big(m_0 + \log(s_0 + \varepsilon)\big)-\big(m_1 + \log(s_1 + \varepsilon)\big)$
          \ENDFOR
          \STATE $\mathrm{LLR} = [\mathrm{LLR} \ \mathbf{l}_{\mathsf{sym}}]$
        \ENDFOR
        \STATE \textbf{Return:} $\mathrm{LLR}$
    \end{algorithmic}
\end{algorithm}

\vspace{-4mm}
\subsection{Shaping Decoder}
\label{subsec:shap_dec}
Let $\mathcal{S}$ denote the set of all constellation points, including both the outer and inner constellation points. The equalized symbols are divided into vectors $\mathbf{y}$ of length $z$. The receiver computes the minimum distance metric:
\begin{align}
    \label{eq:shap_dec1}
    \mathbf{d} = \underset{s\in\mathcal{S}}{\min} \ \Vert \mathbf{y} - s\Vert,
\end{align}
where $\mathbf{d} \in \mathbb{R}_+^{z \times 1}$ is the vector of minimum distances. Let $g$ be a function defined as:
\begin{align}
    \label{eq:shap_dec2}
    e(\mathbf{s}[i]) = \begin{cases}
        1\quad\quad \mathrm{if} \ \mathbf{s}[i] \in \mathcal{S}_{\mathsf{o}} \\
        0\quad\quad \mathrm{if} \ \mathbf{s}[i] \in \mathcal{S}_{\mathsf{i}}
    \end{cases},
\end{align}
where $\mathcal{S}_{\mathsf{o}}$ and $\mathcal{S}_{\mathsf{i}}$ respectively denote the set of points on the outer and inner constellation ($\mathcal{S} = \mathcal{S}_{\mathsf{o}} \cup \mathcal{S}_{\mathsf{i}}$). Define a function $e$ that returns $1$ if the constellation point is picked from outer constellation and $0$ otherwise. The receiver detects each $z$-length information symbols as:
\begin{align}
    \label{eq:shap_dec3}
    \hat{\mathbf{s}}[i] = \underset{s\in \mathcal{S}}{\arg \min} \ \Vert \mathbf{y} - s\Vert.
\end{align}
The shaping codeword is decoded as:
\begin{align}
    \label{eq:shap_dec4}
    \hat{\mathbf{c}} = e(\hat{\mathbf{s}}).
\end{align}
However, since the shaping codeword has a pre-defined structure (for example, sparsity) the decoded codeword may not always be a valid codeword. 
\textcolor{black}{If the estimate $\hat{\mathbf{c}}$ is not a shaping codeword we change one or more entries to produce a shaping codeword. If $\mathbf{d}[i]$ is the distance between the $i$th received symbol and the $i$th symbol of $\hat{\mathbf{c}}$, then larger metrics $\mathbf{d}[i]$ correspond to less reliable symbols. We identify the symbol with largest metric and flip that
symbol (replacing an outer symbol with the paired inner symbol or vice versa).}
If this results in a valid codeword, the process is stopped else the process is continued to find the next dirtiest symbol index and so on. This procedure gives the decoded shaping codeword. 

Wireless systems employ LDPC codes. The LDPC decoder requires log-likelihood ratios (LLRs) for decoding. In the following Subsection we describe how to compute the LLRs for the OFDM system with shaping.
\vspace{-2mm}
\subsection{LDPC Decoding}
\label{subsec:ldpc_dec}
LLR computation for LDPC decoding is presented in Algorithm~\ref{alg:llrs}. The algorithm takes as input the equalized symbols, the constellation points, the label map which corresponds to the labels for both inner and outer constellation (each label is mapped to exactly two symbols), noise variance, and the prior probabilities of the constellation points in the log scale. From Steps 3 -- 9, a minimum distance metric is computed and the prior probabilities are introduced\textcolor{black}{\footnote{\textcolor{black}{The probability of $1$s in the shaping code provides the priors for the outer constellation points and the probability of $0$s provides the priors for inner constellation points. This is because once the choice is made between the inner and outer constellation, the information symbols are chosen uniformly at random.\label{foot:log_prior}}}}. From Steps 10 -- 15, the Algorithm combines the metrics for all the points that have the same label using the log sum exponential approximation for improving numerical stability~\cite{bishop2006pattern}. Combining metrics is essential since more than one constellation point is mapped to the same index (see Sec.~\ref{subsec:shap_enc}). Steps 16 through 22 compute the logits to LLR conversion in the usual way.

\vspace{-2mm}
\subsection{Rate and PAPR}
\label{subsec:rate_papr}
The extra rate introduced by the shaping code is through power modulation. The rate is a function of number of codewords that are present in the shaping code. Let $\mathcal{C}$ denote the set of $z$-length shaping codewords (or the codebook). Then the rate achieved by the shaping codebook is:
\begin{align}
    \label{eq:rate_papr_comp1}
    r = \frac{\log_2\vert\mathcal{C}\vert}{z},
\end{align}
where the division by $z$ indicates that $\log_2\vert\mathcal{C}\vert$ bits are transmitted in $z$ channel uses. This rate is in addition to the rate achieved by the system without shaping.

In practice, PAPR is an important consideration. A high PAPR requires an amplifier with large linear operating region~\cite{nee_prasad_ofdm_2000}, which may not always be realizable in practice. When the constellation is augmented, we increase the peak power of the constellation. The average energy of the augemented constellation is:
\begin{align}
    \label{eq:rate_papr_comp2}
    P_{\mathsf{avg}} = \pi_{\mathcal{S}_{\mathsf{i}}}P_{\mathsf{i}} + \pi_{\mathcal{S}_{\mathsf{o}}}P_{\mathsf{o}},
\end{align}
where $\pi_{\mathcal{S}_{\mathsf{i}}}$ ($\pi_{\mathcal{S}_{\mathsf{o}}}$) denotes the probability of picking a point from the inner (outer) constellation, and $P_{\mathsf{i}}$ ($P_{\mathsf{o}}$) is the average power of the inner (outer) constellation. The PAPR can be written as:
\begin{align}
    \label{eq:rate_papr_comp3}
    \mathrm{PAPR} = \frac{\underset{s\in\mathcal{S}}{\max}\ \vert s \vert^2}{P_{\mathsf{avg}}}.
\end{align}
For the 4-QAM system in Fig.~\ref{fig:const_4_qam}, with $3$-length shaping with sparsity $1$, $P_{\mathsf{avg}} = 0.75\cdot2 + 0.25\cdot 10 = 4$ and the $\mathrm{PAPR} = 2.5$.


\begin{figure}
    \centering
    \includegraphics[width=\linewidth]{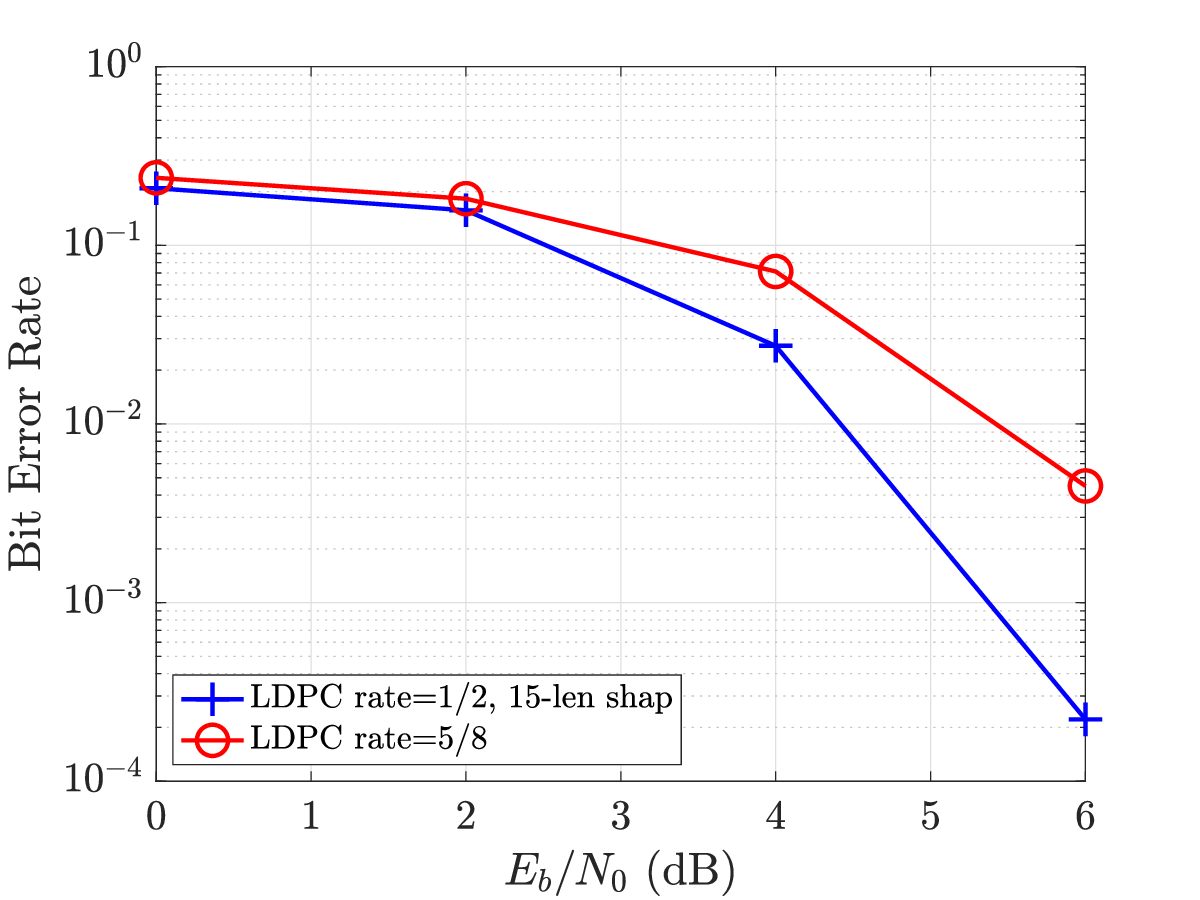}
    \caption{
    The rate $1/2$ LDPC code with a $1$-sparse length $15$ shaping code has the same transmission rate as the rate $5/8$ LDPC code, but improves performance by $1.5$ dB at a BER of $10^{-2}$.}
    \label{fig:4_qam_15_len_shap}
    \vspace{-1mm}
\end{figure}

\section{Numerical Results}
\label{sec:num_results}
In this Section we present the numerical results evaluating the performance of the OFDM scheme with shaping. For all the results presented here, we consider a practical setting where estimated channel is used for equalization along with LDPC encoding. We consider an OFDM system with $M=72, N=14, \Delta f = 30$ kHz. For all the simulations we consider the Vehicular-A channel model~\cite{veh_a} with a maximum Doppler spread $\nu_{\max}$ of $815$ Hz. Each path Doppler is obtained using the Jake's spectrum $\nu_p = \nu_{\max}\cos(\theta)$, where $\theta\sim\mathsf{U}[-\pi, \pi)$. \textcolor{black}{We use the 5G LDPC encoder and decoder provided by Sionna~\cite{sionna}. The message length $k$ and code length $n$ are specified for each code in the following subsections.}

\vspace{-2mm}
\subsection{$4$-QAM, $15$-length shaping}
\label{subsec:4_qam_15_shap}
In this Subsection, we consider a $15$-length shaping code with sparsity $1$. We have $16$ codewords in the shaping codebook which convey $4$ bits of information over $15$ channel uses. The additional rate is $\approx 0.25$. We pair this with an LDPC code of rate $0.5$ \textcolor{black}{($k=855, n=1710$)}, and the effective rate is $1.25$. For the unshaped performance we consider a rate $5/8$ \textcolor{black}{($k=855, n=1710$)} LDPC code with $4$-QAM that also achieves rate $1.25$. Figure~\ref{fig:4_qam_15_len_shap} compares the bit-error performance of the two systems. It is seen that the performance of the OFDM system with shaping is superior compared to that without shaping and the performance gap increases with increase in signal power. Shaping gain~\cite{calderbank2002nonequiprobable} and lower rate for LDPC with shaping both contribute to this performance improvement.
\vspace{-2mm}
\subsection{$4$-QAM, $23$-length shaping}
\label{subsec:4_qam_23_shap}
\begin{figure}
    \centering
    \includegraphics[width=\linewidth]{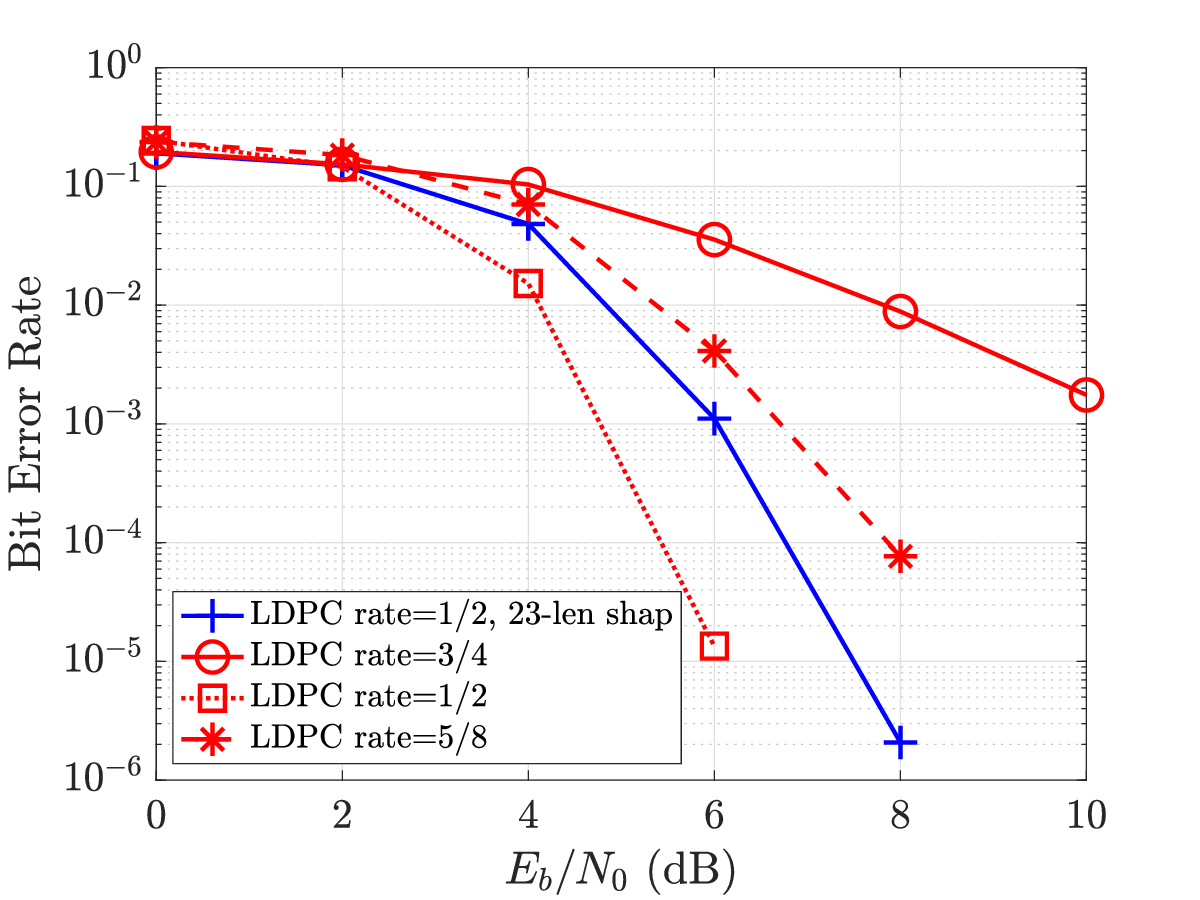}
    \caption{
    The rate $1/2$ LDPC code with a $3$-sparse length $23$ shaping code has the same transmission rate as the $3/4$ LDPC code, but improves performance by $4$ dB at a BER of $10^{-3}$. It has slightly better BER performance than the rate $5/8$ LDPC code but improves transmission rate by $20$\%.}
    \label{fig:4_qam_23_len_shap}
    \vspace{-1mm}
\end{figure}

In this Subsection, we consider a $23$-length shaping code with sparsity $3$. We have $2048$ codewords in the shaping codebook which convey $11$ bits of information over $23$ channel uses. The additional rate is $\approx 0.5$. We pair this with an LDPC code of rate $0.5$ \textcolor{black}{($k=851, n=1702$)}, and the effective rate is $1.5$. For the unshaped performance we consider a rate $3/4$ \textcolor{black}{($k=1296, n=1728$)} LDPC code with $4$-QAM that also achieves rate $1.5$. Figure~\ref{fig:4_qam_23_len_shap} shows the bit-error performance of the shaped and unshaped OFDM systems. The performance gap is wider here, because to match the rate, the unshaped OFDM system uses a higher rate LDPC, which is less capable of correcting errors.
\vspace{-2mm}
\subsection{$16$-QAM, $23$-length shaping}
\label{subsec:16_qam_23_shap}
\begin{figure}
    \centering
    \includegraphics[width=\linewidth]{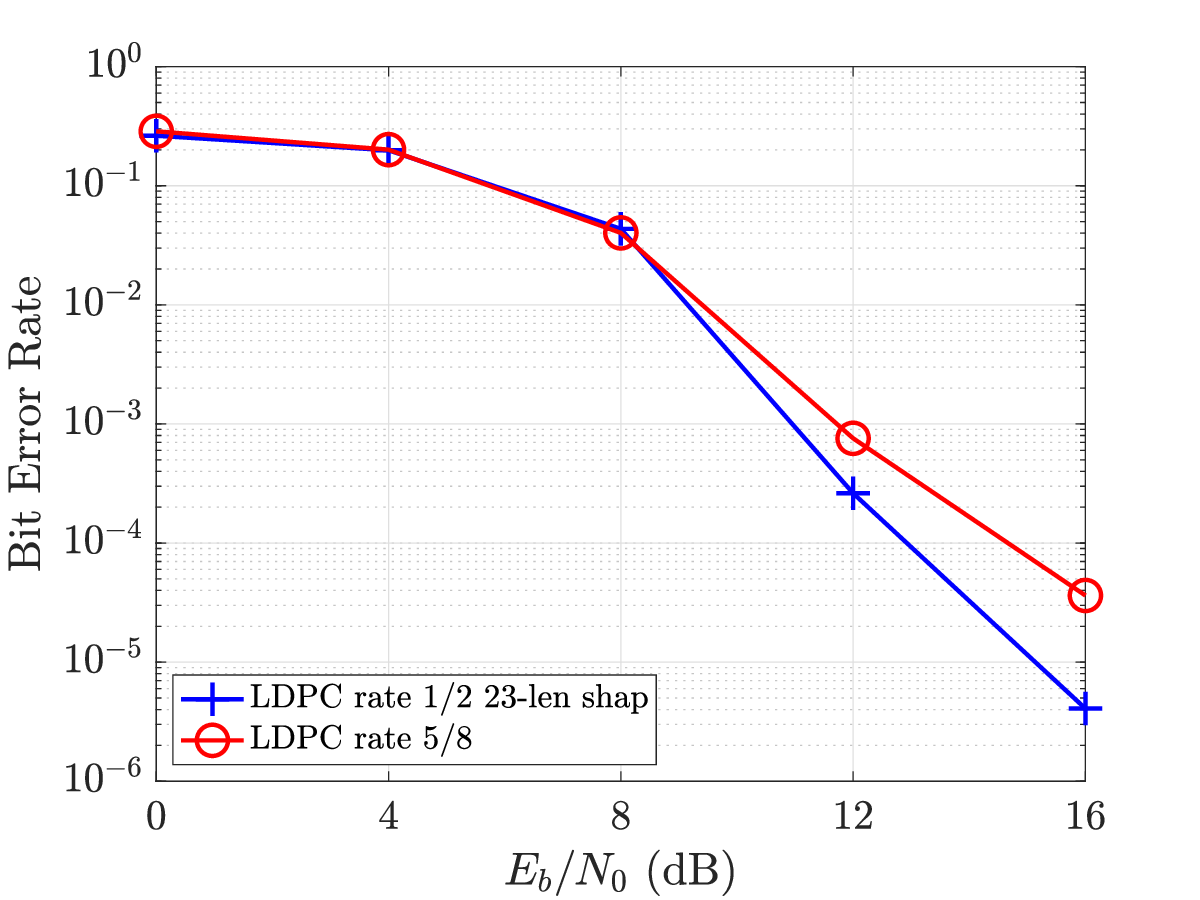}
    \caption{Coded bit-error performance comparison between OFDM modulation with and without shaping. $16$-QAM, $23$-length shaping with $3$ sparsity.}
    \label{fig:16_qam_23_len_shap}
    \vspace{-1mm}
\end{figure}

In this Subsection, we consider the same $23$-length shaping code with LDPC code of rate $0.5$ \textcolor{black}{($k=1702, n=3404$)}. The effective rate is $2.5$ since we use $16$-QAM. For the unshaped performance we consider a rate $5/8$ \textcolor{black}{($k=2160, n=3456$)} LDPC code with $16$-QAM that also achieves rate $2.5$. Figure~\ref{fig:16_qam_23_len_shap} shows the bit-error comparison. At high signal energies there is performance gain of about $1$ dB over the unshaped OFDM system.

\begin{figure}
    \centering
    \includegraphics[width=\linewidth]{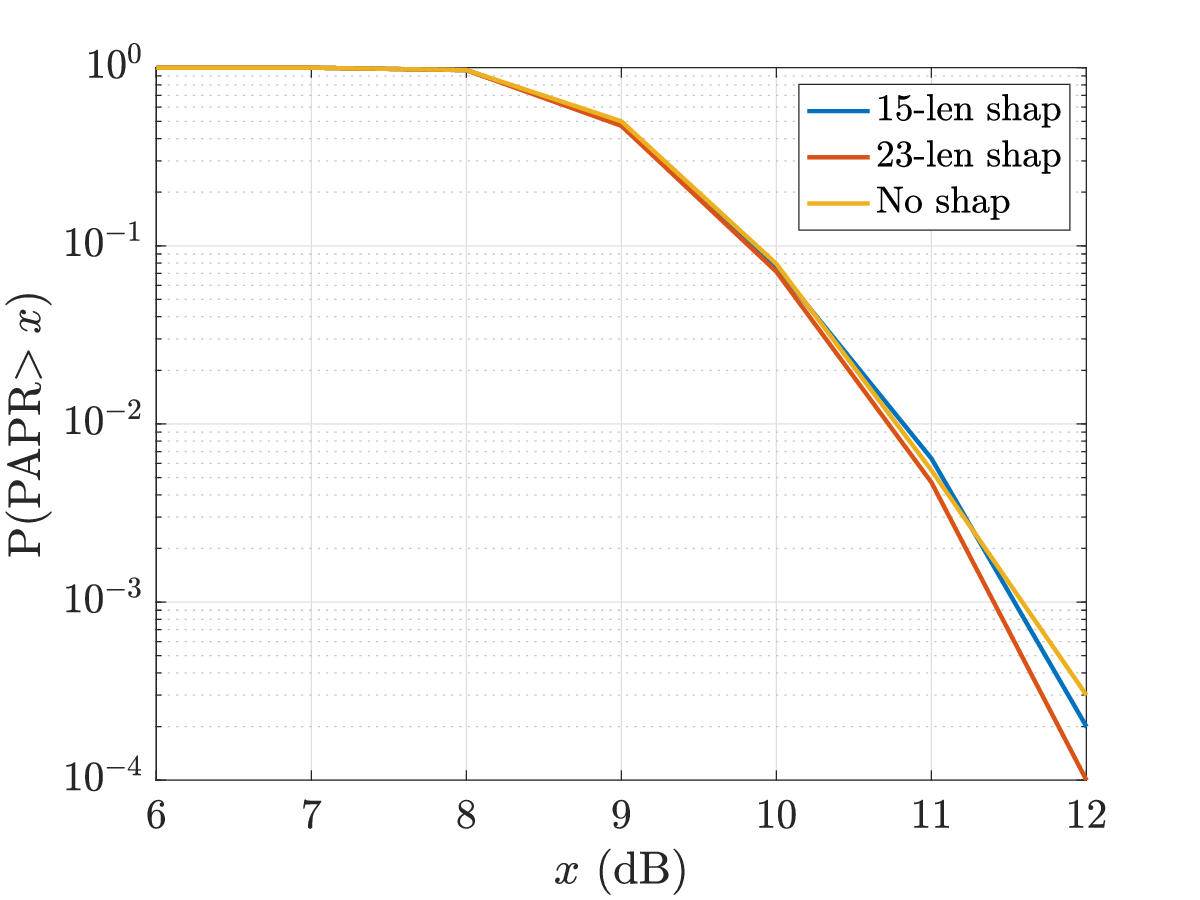}
    \caption{\textcolor{black}{CCDF of PAPR comparing the performance of OFDM with no shaping, $15$-length shaping and $23$-length shaping for 4-QAM constellation.}}
    \label{fig:papr_4_qam}
    \vspace{-1mm}
\end{figure}

\vspace{-2mm}
\subsection{\textcolor{black}{PAPR}}
\label{subsec:papr}
\textcolor{black}{Figure~\ref{fig:papr_4_qam} shows the PAPR of the OFDM system with and without shaping. 4-QAM constellation is used for communication. It is seen that although the PAPR of the OFDM system is expected to increase when the information symbols are picked from the outer constellation, after the IDFT block at the transmitter, the difference in PAPR between the systems with and without shaping is negligible. This is because of the sparsity of the shaping codeword and also the fact that DFT bases spread the energy over all time domain symbols.}


\section{Conclusions}
\label{sec:conclusions}
We aimed to improve performance at the cell edge, where the $5$G wireless standard specifies a suite of LDPC codes with different rates that are applied to $4$-QAM. We have demonstrated that the combination of rate $1/2$ LDPC coding and shaping improves upon the performance of this baseline suite of codes. The rate $1/2$ LDPC code with a $3$-sparse length $23$ shaping code has the same transmission rate as the rate $3/4$ LDPC code, but improves performance by $4$ dB at a BER of $10^{-3}$. The same combination has slightly better BER performance than the rate $5/8$ LDPC code but improves transmission rate by $20$\%. Our results suggest that at the cell edge where transmission rates are low, there is significant value in using shaping to contribute transmission rate. When decoding the combination of LDPC codes and shaping codes, we have described a method of calculating LLRs that may be of independent interest.


\bibliographystyle{IEEEtran}
\bibliography{references}

\end{document}